\documentstyle[preprint,aps]{revtex}
\begin{document}
\pagestyle{plain}
\draft
\preprint{CCAST-95-010}
\title{Different Nuclear Dependence of $J/\psi$ and
$\psi^{\prime}$ Production \\
in p-A and A-p Collisions
	\footnote{\it Partly supported by the National Natural
		   Science Foundation of China}}

\author{Yong-Bin He$^{1,2}$
\footnote{E-mail: hyb@ccastb.ccast.ac.cn, Telefax: (086)-010-2562586},
Wei-Qin Chao$^{1,3,4}$ and Chong-Shou Gao$^{1,2,3}$}
\address{ 1. China Center of Advanced Science and Technology (World
Laboratory),
        \\P.O. Box 8730, Beijing 100080,  China\\
	2. Physics Department, Peking University,
         Beijing 100871, China\\
	3. Institute of Theoretical Physics, Academia Sinica, P.O. Box
	2735, Beijing 100080, China\\
	4. Institute of High Energy Physics, Academia Sinica, P.O. Box
	918(4),	Beijing 100039, China\\}
\date{\today}
\maketitle

\begin{abstract}
While presently available p-A data
observe that the nuclear dependence
of $J/\psi$ and $\psi^{\prime}$ production
is the same within errors, we show
that different nuclear dependence of
$J/\psi$ and $\psi^{\prime}$ production
in the kinematic region uncovered by the present p-A data
can be predicted based on two different nuclear absorption
scenarios. It is found that
the predicted production ratio $\sigma (\psi^{\prime})/\sigma (J/\psi)$
at positive $x_F$ in A-p collisions
is mainly determined by nuclear absorption, and hence allows
direct experimental test of nuclear absorption scenarios.
\end{abstract}

\narrowtext
\vfill

\newpage
Quarkonium production in
experiments
with a proton-beam incident on a nucleus target,
{\it i.e.}~{\em p-A collisions},
has attracted much attention
in recent years in connection with $J/\psi$
suppressions in high energy heavy-ion collisions
\cite{mat86,abr88,gro89,bag89}.
Actually it is a good place to
check the
hadronic absorption models which attribute the observed $J/\psi$
suppressions in nucleus-nucleus collisions to the absorption of $J/\psi$
through its interactions with hadronic environment instead of the formation
of quark-gluon plasma (QGP).
In particular, the experimental observations
\cite{ald91,abr94} that
the $J/\psi$ and $\psi^{\prime}$ production
has the same nuclear dependence in p-A collisions
put a strict constraint on various absorption models.
However, it has been realized that
all present experiments for the production of $J/\psi$ and $\psi^{\prime}$
in p-A collisions
studied only $J/\psi$ or $\psi^{\prime}$ which is fast
in the lab system or has positive Feynman $x_F$.
In fact, in the kinematic region not yet probed by
present p-A experiments, {\it i.e.} the
negative $x_F$ region,
different nuclear dependence of $J/\psi$ and $\psi^{\prime}$
production can be predicted based on different scenarios of
absorption by target nucleons, which will be shown in this work.
One of these scenarios, the
``traditional'' nuclear absorption model (we shall refer to it
as {\em Scenario-I} throughout this work), has been established
in the literature \cite{bro88,bla89,gav90,vog91}
to account
for the effects of absorption by nucleons in the
following way: while the $c\bar c$ pair (pre-meson), produced
as a colour singlet and of small spatial size, is expanding
gradually to its
bound state size, it can be dissociated by nucleons
through an interaction cross section depending on the size of
the $c\bar c$ pair. Very recently another scenario ({\em Scenario-II})
\cite{kha95b},
stemming from the most recent theoretical progress made in hadronic
production of quarkonium states,
has been proposed to explain the observed charmonium suppression
in hadron-nucleus and nucleus-nucleus collisions. This scenario
considers charmonium production through the intermediate
next-to-leading Fock space state consisting of a colour octet
$c\bar c$ pair and a soft gluon. The interaction between this composite
state and nucleons can lead to charmonium suppression.
In this work we shall show how these  two different
absorption scenarios could accommodate
present p-A data and yet predict different nuclear dependence of
$J/\psi$ and $\psi^{\prime}$ production in p-A collisions
at negative $x_F$.

At negative $x_F$ in p-A collisions
the mechanisms of charmonium production
such as nuclear shadowing \cite{qiu87,ash88,bro90}
and intrinsic charm \cite{bro89,vog91}
are expected to be
unimportant and will be neglected in our work.
Initial state energy loss effects
\cite{gav92} may lead to an increase of
charmonium production in the negative $x_F$ region,
which gives an opposite trend to the experimental data.
Furthermore, the usual EMC
effects \cite{aub83} can be important at negative $x_F$ \cite{kha95a}.
The initial state energy loss and EMC effects are not included in our study
and their effects will be discussed later.
Our previous work \cite{cha95}
showed that $x_F$-dependent
comover contributions, as well as the nuclear absorptions,
are essential
for a comparison between theories and experimental data at small $x_F$.
We shall include only the effects of absorption by target nucleons and
comovers in our study of
$J/\psi$ and $\psi^{\prime}$ production in
p-A collisions at negative $x_F$.
In fact, recent results
on open charm production near $x_F$=0 \cite{lei94}
suggested  that absorptions may be the dominant cause of the
charmonium suppression in p-A collisions at negative $x_F$.

We shall show in this work that although
the individual $J/\psi$
and $\psi^{\prime}$ production at negative $x_F$ in p-A collisions
predicted by the two different absorption scenarios may
depend on the choice of parameters in  comover contributions,
the ratio of the two production cross sections
$\sigma (J/\psi)/\sigma (\psi^{\prime})$ is
insensitive to comover contributions, and
therefore the  prediction
of the ratio may allow an experimental
test of nuclear absorption scenarios.

However, a study of charmonium production at negative $x_F$
(corresponding to the case of
charmonium moving slow in the rest frame of target nucleus)
is essentially
impossible in an experiment of proton-beam incident on a nucleus target
(p-A collisions).
An alternative is the experiment of
a heavy ion beam incident on a hydrogen
or deuterium target
\cite{kha95a,hoy95}, which we shall call
{\em A-p collisions}.
In A-p collisions the produced charmonia are moving fast in the lab system
and thus the relevant measurements are more readily.
In this work we have also
calculated and compared
the predictions of the mentioned two absorption
scenarios for $J/\psi$ and $\psi^{\prime}$
production in A-p collisions at CERN-SPS energy.

In Scenario-I for p-A collisions
the produced charmonium states ($J/\psi$ or $\psi^{\prime}$)
or $c\bar c$ pairs (pre-mesons) can been dissociated by nucleons
in the target nucleus into open-charm pairs ($D \bar D$), which leads to
suppressions of charmonium production. However, the formation of charmonium
bound state requires finite proper time, and this formation time
will suffer a Lorentz delay in the lab system.
In the kinematic region reached
by present experiments, the produced $c\bar c$
pair is moving fast in the lab system and
may have been outside the target nucleus
before it forms a physical charmonium bound state
due to the Lorentz delay of the formation time in the lab system. Therefore,
present experimental data show that the $J/\psi$ and $\psi^{\prime}$
productions have the same nuclear dependence in p-A collisions,
although the radii of
the $J/\psi$ and $\psi^{\prime}$
bound states differ by almost a factor of two in potential
models \cite{kwo87}.
According to this picture, those $c\bar c$ pairs
which are slow in the lab system will have enough time to form the
charmonium bound states ($J/\psi$ or $\psi^{\prime}$,
 {\it etc.}) inside the target
nucleus, and thus the interaction of fully formed bound states $J/\psi$
and $\psi^{\prime}$ with the surrounding target nucleons
may lead to different nuclear dependence for
$J/\psi$ and $\psi^{\prime}$
production through the nuclear absorption depending
on the final-state size, as we shall see below by explicit calculations.

Including only the effects of absorption by nucleons and comovers,
the $x_F$-dependent
cross section for quarkonium production in
p-A collisions is
\begin{eqnarray}
\label{crosseq}
\nonumber
{d\sigma^{pA} \over dx_F} = A {d\sigma^{pp} \over dx_F}\int d^2 b
\int\limits_{-\infty}^{+\infty}dz
\rho_A(b,z)exp\left\{-\int\limits_z^{+\infty}
dz^{\prime}\rho_A(b,z^{\prime})\sigma_{abs}(z^{\prime}-z)\right\}
\\
\times
exp\left\{-\int_{\tau_0}^{\tau_f}d\tau \sigma_{co} v_{re} n(\tau,b)\right\},
\end{eqnarray}
where $\sigma^{pp}$ is the bare nucleon-nucleon cross section,
$b$ the impact parameter and $\rho_A(b,z)$ the nuclear density.

The second exponential factor in Eq.~(\ref{crosseq})
represents the comover contributions,
where  $\sigma_{co}$ is the ($c \bar c$)-comover
absorption cross section, $\tau_0$ is the formation time of comovers,
$\tau_f$ is the effective proper time over which the comovers can interact
with $c \bar c$ pair, $v_{re}$ is the relative velocity of $c \bar c$
with the comovers, and
$n(\tau,b)$ is the density of comovers at the proper time $\tau$ and
impact parameter $b$. In our calculations we have introduced the
$x_F$ dependence of comover contributions
through $n(\tau,b)$ as we did in Ref.~\cite{cha95},
where $n(\tau,b)$ is related to the $x_F$-dependent comover distribution
that is derived from the rapidity distribution of comovers.
We fix $\tau_f=r_0/c_s$ with $r_0\simeq$1.2~fm
and $c_s\simeq 1/\sqrt{3}$, $v_{re} \simeq$0.6
\cite{vog91}. We shall vary
the values
of $\tau_0$ and $\sigma_{co}$ to show the influence of comover contributions
on the charmonium production at negative $x_F$.

The first exponential in Eq.~(\ref{crosseq})
is the survival probability of the $c\bar c$
pair after its interaction with target nucleons.
In the context of Scenario-I the nuclear absorption cross section is
assumed to depend on the spatial separation of
the $c\bar c$ pair, so that
\begin{eqnarray}
\label{abs1}
\sigma_{abs}(z^{\prime}-z)=
\left\{
\begin{array}{ll}
\sigma_{RN}(\tau /\tau_{R})^k &~~~~if~\tau < \tau_{R},
\\
\sigma_{RN} &~~~~ otherwise,
\end{array}
\right.
\end{eqnarray}
where in general $k\agt$0 and we shall take $k$=2 in our calculations.
Here $R$ stands for charmonium bound states $J/\psi$ or $\psi^{\prime}$.
The formation time of $\psi^{\prime}$
is taken as $\tau_{\psi^{\prime}}\simeq$1.5 ~fm \cite{kar88}, and
the effective formation time of $J/\psi$ bound state is estimated to be
$\tau_{\psi} \simeq$1.2 ~fm, taking into account the fact that about
70\% of the observed $J/\psi$ are
directly produced and the remaining 30\% come from the decay of
$\chi_c$ \cite{gav94}.
We shall take the nuclear absorption cross sections for $J/\psi$
and $\psi^{\prime}$ as $\sigma_{\psi N}$=4~mb from
geometric considerations and
$\sigma_{\psi^{\prime} N}=\sigma_{\psi N}(r_{\psi^{\prime}}/r_{\psi})^2
\simeq 3.7\sigma_{\psi N}$ \cite{pov87,gav94}.

The proper time $\tau$
in Eq. (\ref{abs1}) is related to the distance traveled by
the $c\bar c$ pair through
$\tau=(z^{\prime}-z)/\gamma v$, $v$ is the velocity
of the $ c \bar c$ pair in the rest frame of target nucleus, and
$\gamma =(1-v^2)^{-1/2}$.
We can relate the velocity of the $c\bar c$ pair
in the rest frame of target nucleus
to the Feynman $x_F$ by
\begin{eqnarray}
\label{gamma}
\gamma v={P \over M}=\gamma_{cm}(P_L^*+\beta_{cm}\sqrt{(P_L^*)^2+M^2}),
\end{eqnarray}
where $P_L^{*}=x_F (P_L^*)_{max}$ is the longitudinal momentum of the $c\bar c$
pair in the center-of-mass system and its maximum is
\begin{eqnarray}
(P_L^{*})_{max}=\sqrt{[s-(M+2m_p)^2][s-(M-2m_p)^2]}/2\sqrt{s}.
\end{eqnarray}
Here $M$ is the mass of the $c\bar c$ system, $m_p$ the mass of a proton,
and $s$ the center-of-mass energy squared.
The transformation from the rest frame of the target nucleus
to the center-of-mass system is given by
$\gamma_{cm}=(E_b+m_p)/\sqrt{s}$, and
$\beta_{cm}=P_b/(E_b+m_p)$, where $E_b$ and $P_b$ are the energy and momentum
of the projectile beam in the lab system.
One can see from Eq.~(\ref{gamma}) that a $J/\psi$ produced at
negative $x_F$ corresponds
to a slow $J/\psi$ in the rest frame of the target nucleus.

We have calculated the $x_F$ dependence of $\alpha$
for $J/\psi$
and $\psi^{\prime}$ production at negative $x_F$
in Fermilab E772/E789 p-A collisions
in the parametrization,
\begin{eqnarray}
\label{para}
{d\sigma^{pA} \over dx_F} = {d\sigma^{pp} \over dx_F}A^{\alpha(x_F)}.
\end{eqnarray}
The results
are shown in Fig.~1
by the upper two curves, with the solid curve for $J/\psi$
production and the dashed for $\psi^{\prime}$ production,
parameters in comover contributions taken as $\sigma_{co}$=4~mb,
$\tau_0$=0.8 ~fm.
Also shown are the Fermilab
E789 data on the $J/\psi$ production at $x_F\alt$0 \cite{lei92}.
The curves are
a little above the E789 data because for $J/\psi$ production
the nuclear shadowing effect can
contribute somehow at $x_F\simeq$0 \cite{cha95}.
It can be seen that Scenario-I actually gives the same nuclear dependence
for $J/\psi$ and $\psi^{\prime}$ production in p-A collisions at
$x_F\agt$0,
as E772/E789 experiments have observed.
However, different nuclear dependence for $J/\psi$
and $\psi^{\prime}$ production
appears at $x_F \alt -0.2$, which is beyond the E772/E789
data. Note that
the rise of $\alpha_{J/\psi}$ (solid curve in Fig.~1) at negative $x_F$
is due to the $x_F$-dependent comover
contributions in Eq.~(\ref{crosseq}).

As we discussed above, in order to confront experiments we have
calculated the production of $J/\psi$ and $\psi^{\prime}$
in A-p collisions at CERN-SPS energy $\sqrt{s}$=17.4~GeV.
As a matter of fact,
the production at positive $x_F$ in A-p collisions
corresponds to the production at negative $x_F$ in p-A collisions.
In Fig.~2 we show the results of $J/\psi$ and $\psi^{\prime}$
production in A-p collisions at CERN-SPS energy
for two sets of parameters $\sigma_{co}$
and $\tau_0$ in comover contributions:
the upper two curves with
$\sigma_{co}$=2~mb and $\tau_0$=1.0 ~fm
(dot-dashed curve for $\alpha_{J/\psi}$ and
dotted curve for $\alpha_{\psi^{\prime}}$);
the lower two curves with $\sigma_{co}$=4~mb and $\tau_0$=0.8 ~fm
(solid curve for $\alpha_{J/\psi}$ and
dashed curve for $\alpha_{\psi^{\prime}}$).
One sees that comover contributions are essential for
charmonium production at positive $x_F$
in A-p collisions.

In order to reduce the uncertainties caused by the comover contributions,
we have studied the production ratio
$R_{\psi^{\prime}/\psi} \equiv \sigma (\psi^{\prime})/f\sigma (J/\psi)$
in A-p collisions at CENR-SPS energy,
keeping in mind that comover contributions depends weakly on
the type of produced charmonium. Here $f\simeq$ 0.14 is the
experimental $\psi^{\prime}$ and $J/\psi$ production ratio
in hadron-hadron collisions \cite{abr94,gava94}.
The results are illustrated in
Fig.~3 by the upper two curves: the solid curve for parameters
$\sigma_{co}$=4~mb, $\tau_0$=0.8~fm, and the dashed curve
for $\sigma_{co}$=2~mb, $\tau_0$=1.0~fm. We see that the production
ratio is insensitive to comover contributions.
Moreover, in addition to
nuclear absorption and comover contributions the EMC effects can
be important at positive $x_F$
in A-p collisions \cite{kha95a}.
However, since the EMC effects are initial-state
effects and should be independent of the final produced type of charmonium,
the inclusion of the EMC effects will not affect the value
of the production ratio $R_{\psi^{\prime}/\psi}$.
The same argument exists for initial-state energy loss \cite{gav92}.
Therefore we suggest that the prediction of
production ratio $R_{\psi^{\prime}/\psi}$
at positive $x_F$ in A-p collisions
should be able to provide a direct test for
nuclear absorption scenario discussed above.

Another nuclear absorption scenario (Scenario-II)
has been recently proposed by Kharzeev and Satz \cite{kha95b},
and shown to be able to derive the
phenomenological Gerschel-H\"{u}fner fit
\cite{ger92} for hadron-nucleus data on
$J/\psi$ production.  In this
scenario charmonium state is considered to be
produced through the intermediate next-to-leading
Fock space component $|(c \bar c)_8g>$ formed by a colour octet $c\bar c$
pair and a gluon, whose lifetime is estimated to be
$\tau_8 \simeq$0.25~fm.
The interaction of the composite state
$(c \bar c g)$ with nucleons will lead to suppression of charmonium
production. In particular, it is argued that
the composite state $(c \bar c g)$ is of the same size for
all charmonium states, and hence the $(c \bar c g)$-nucleon cross section
is the same for all charmonia. As a result, the production of $J/\psi$
and $\psi^{\prime}$ will have the same suppression,
as long as charmonium suppression
is dominated by the interaction of the intermediate $(c \bar c g)$ with
nucleons. This is exactly what happens in the present p-A experiments.

However, this scenario may also
predict different nuclear dependence of
$J/\psi$ and $\psi^{\prime}$ production
 in p-A collisions in the kinematic region where
the composite state $(c \bar c g)$ has turned into the fully formed
charmonium bound states $J/\psi$ and $\psi^{\prime}$,
and then nuclear absorption will depend on the specific
type of charmonium state. This will be shown in the following calculations.

Different from Scenario-I the nuclear
absorption cross section in Scenario-II is now assumed to be
\begin{eqnarray}
\label{abs2}
\sigma_{abs}(z^{\prime}-z)=
\left\{
\begin{array}{ll}
\sigma_{(c\bar c g)N} &~~~~if~z^{\prime}-z < \gamma v \tau_8,
\\
\sigma_{RN} &~~~~ otherwise,
\end{array}
\right.
\end{eqnarray}
where $\gamma v$ is given by Eq.~(\ref{gamma}), $\sigma_{RN}$ are the relevant
nuclear absorption cross sections as the same as those in Eq.~(\ref{abs1}),
$\tau_8\simeq$0.25~fm is the lifetime of the intermediate state
$(c \bar c g)$ \cite{kha93,kha95b}.
The cross section of the interaction between the
intermediate state $(c \bar c g)$ and nucleons $\sigma_{(c\bar c g)N}$
is estimated to be 6~-~7~mb \cite{kha95b}. We shall take
$\sigma_{(c\bar c g)N}$=7~mb in our calculations.

We have also in Scenario-II completed a calculation of
the $x_F$ dependence of $\alpha$ in the parameterization Eq.~(\ref{para})
for $J/\psi$
and $\psi^{\prime}$ production in p-A collisions at Fermilab
E772/E789 energy. The results for chosen parameters,
$\sigma_{co}$=4~mb  and $\tau_0$=0.8~fm,
are shown in Fig.~1 by the lower two
curves: dot-dashed curve for $J/\psi$ production
and dotted curve for $\psi^{\prime}$ production. One sees that
at $x_F\agt$0, {\it i.e.} in
the kinematic region covered by E772 data on $J/\psi$
and $\psi^{\prime}$ production, Scenario-II also gives nearly
the same nuclear dependence of
$J/\psi$ and $\psi^{\prime}$
production. On the other hand, a rather different nuclear
dependence of $J/\psi$ and $\psi^{\prime}$ production
is predicted by Scenario-II at $x_F\alt -0.1$. We can attribute this difference
to the nuclear absorption dominated by the interaction of the fully
formed $J/\psi$ or $\psi^{\prime}$ bound states  with target nucleons.
However, the predicted $J/\psi$ production at $x_F\simeq$0 is more depleted
than the E789 data, which might imply our  over-estimate
of the comover contributions
in this region. The over-estimate, fortunately, will affect little
the production ratio $R_{\psi^{\prime}/\psi}$, as it is shown below.

In Fig.~3 we show
the calculated production ratio
$R_{\psi^{\prime}/\psi}$ in  A-p collisions at CERN-SPS
energy based on Scenario-II by the lower two curves:
dot-dashed curve with $\sigma_{co}$=4~mb, $\tau_0$=0.8~fm;
and dotted curve with $\sigma_{co}$=2~mb, $\tau_0$=1.0~fm. Again one find
that the production ratio $R_{\psi^{\prime}/\psi}$
given by  Scenario-II is insensitive to comover
contributions.
Furthermore, the behavior of $R_{\psi^{\prime}/\psi}$  given by Scenario-II
appears
quite different from that predicted by Scenario-I, which is due to the
following essential differences between Scenario-I and Scenario-II: First,
in order that the $J/\psi$ and $\psi^{\prime}$ production be equally suppressed
at positive $x_F$ in p-A collisions, Scenario-I permits very little nuclear
absorption cross section at $x_F\agt$0 for both $J/\psi$ and $\psi^{\prime}$
production, while in Scenario-II the interaction cross section between the
intermediate state  $(c\bar c g)$ and nucleons is rather large but
the same for $J/\psi$ and $\psi^{\prime}$ production; Second, the proper
formation time of physical $J/\psi$ and $\psi^{\prime}$,
$\tau_{\psi}\simeq$1.2~fm and $\tau_{\psi^{\prime}}\simeq$1.5~fm, is rather
long compared with the proper lifetime of the intermediate state
$(c\bar c g)$, $\tau_8\simeq$0.25~fm, therefore the
stronger suppression of  $\psi^{\prime}$ production
relative to $J/\psi$ production in A-p collisions
given by Scenario-II
happens at lower value of $x_F$ than that given by
Scenario-I (see Fig.~3).
Based on the behavior of predicted $R_{\psi^{\prime}/\psi}$,
an experimental test of nuclear absorption scenarios is possible.

Although uncertainties of the parameters in nuclear absorption scenarios
might reduce the power of prediction, some qualitative behaviors
discussed here
should be maintained by future experimental data. If not, {\it e.g.}
the nuclear dependence of $J/\psi$ and $\psi^{\prime}$ production
remains the same at positive $x_F$ in A-p collisions, some nuclear
absorption scenarios will be excluded as the explanation of quarkonium
suppression observed in nuclear collisions.

To summary, we have shown in this work that
nuclear absorption scenarios are able to predict
different nuclear dependence
of $J/\psi$ and $\psi^{\prime}$ production
at negative $x_F$ in p-A collisions
or equivalently at positive $x_F$ in A-p collisions, although the present
p-A data at $x_F\agt$0 observed the same nuclear dependence
of $J/\psi$ and $\psi^{\prime}$ production \cite{ald91,abr94}.
Furthermore, we find that
the production ratio $\sigma (\psi^{\prime})/\sigma (J/\psi)$
at positive $x_F$ in A-p collisions
is mainly determined by nuclear absorption, but insensitive to
comover contributions, initial-state energy loss and EMC effects,
which can also contribute to charmonium production in this
kinematic region.
Therefore the predictions of
production ratio at positive $x_F$ in A-p collisions
are suggested to be able to provide a rather direct
and experimentally accessible test for
nuclear absorption scenarios.

\newpage
\begin{center}
{\bf FIGURE CAPTIONS}
\end{center}

Fig.~1. Predictions for $J/\psi$ and $\psi^{\prime}$
production in the Fermilab E772/E789
experiment with 800~GeV/c proton-beam incident on nuclear
targets:
the upper two curves given by Scenario-I (solid curve for $J/\psi$ and
dashed curve for $\psi^{\prime}$)
the lower two curves given by Scenario-II
(dot-dashed curve for $J/\psi$ and dotted curve for $\psi^{\prime}$).
Also shown are the Fermilab
E789 data on the $J/\psi$ production \cite{lei92}.

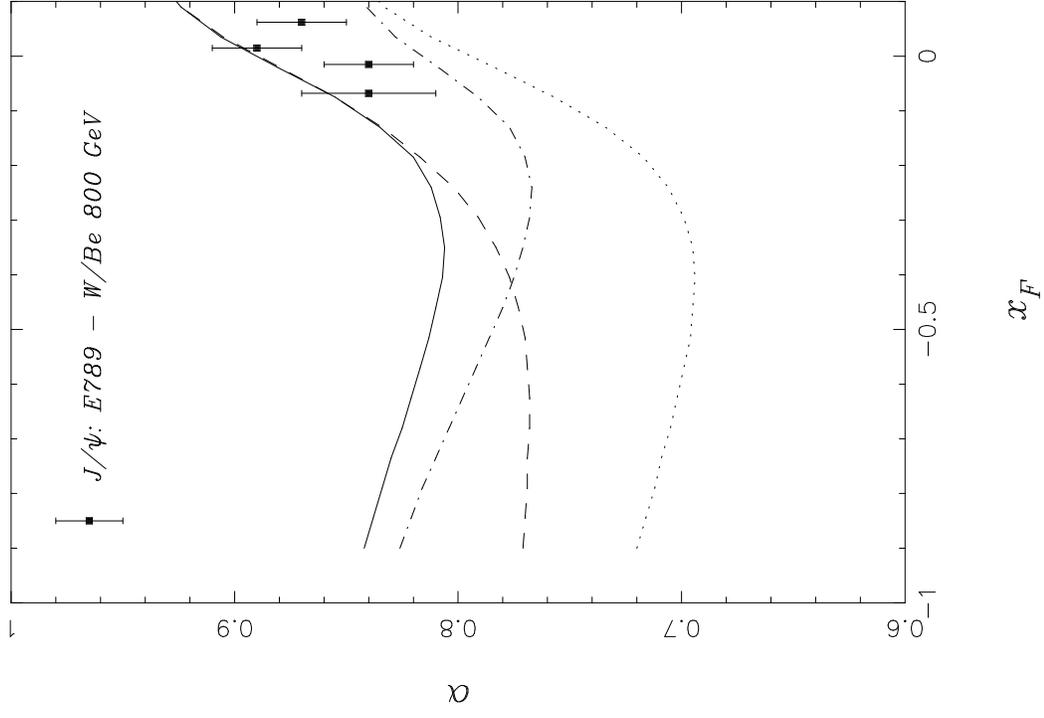
Fig.~2. Predictions given by Scenario-I
for $J/\psi$ and $\psi^{\prime}$
production in the CERN-SPS experiment with 160~GeV/c
$Pb$-beam incident on a hydrogen
target for two sets of parameters $\sigma_{co}$
and $\tau_0$ in comover contributions:
the upper two curves with
$\sigma_{co}$=2~mb and $\tau_0$=1.0~fm
(dot-dashed curve for $\alpha_{J/\psi}$ and
dotted curve for $\alpha_{\psi^{\prime}}$);
the lower two curves with $\sigma_{co}$=4~mb and $\tau_0$=0.8~fm
(dot-dashed curve for $\alpha_{J/\psi}$ and
dotted curve for $\alpha_{\psi^{\prime}}$).

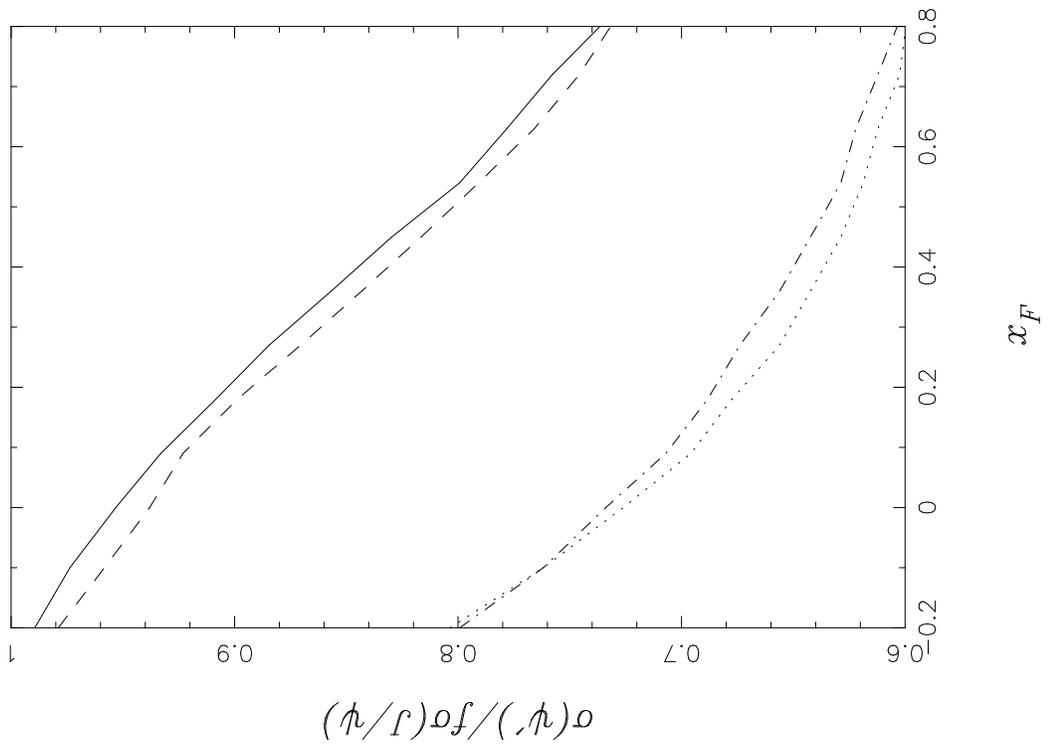
Fig.~3. Predictions of the production ratio for $J/\psi$
and $\psi^{\prime}$ production in the CERN-SPS experiment with 160~GeV/c
$Pb$-beam incident on a hydrogen target given by
Scenario-I (the upper two curves: the solid curve for parameters
$\sigma_{co}$=4~mb, $\tau_0$=0.8~fm, and the dashed curve
for $\sigma_{co}$=2~mb, $\tau_0$=1.0~fm)
and Scenario-II (the lower two curves:
dot-dashed curve with $\sigma_{co}$=4~mb, $\tau_0$=0.8~fm;
and dotted curve with $\sigma_{co}$=2~mb, $\tau_0$=1.0~fm).
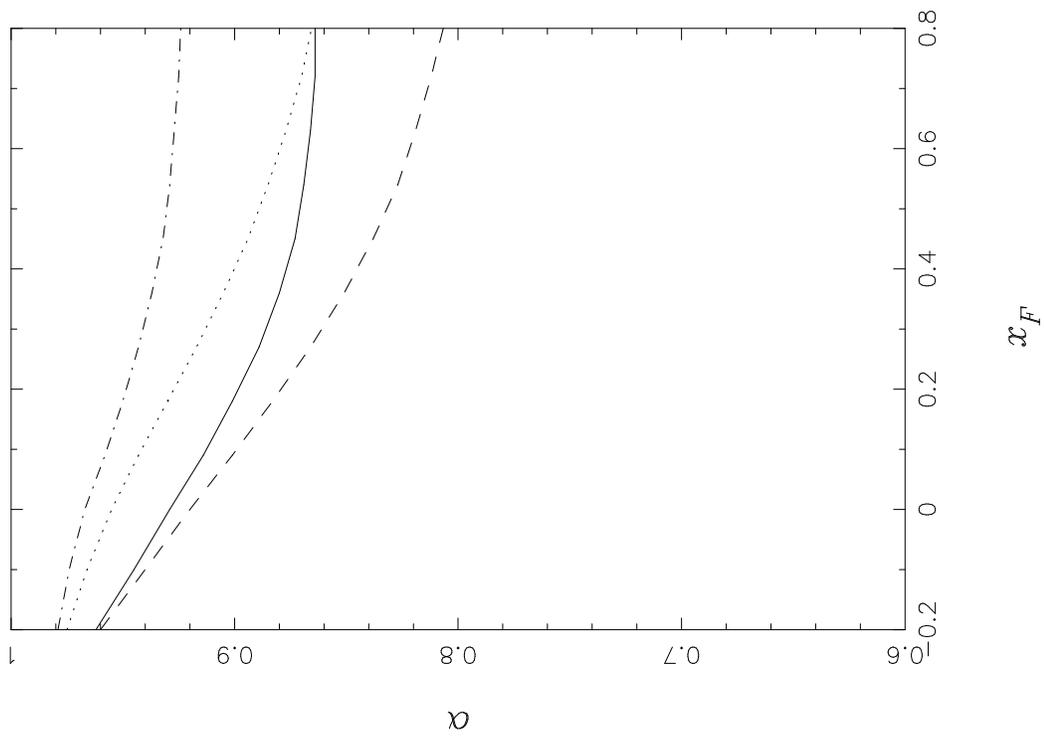
\end{document}